# Ethanol and Hydrogen gas-sensing properties of CuO–CuFe$_2$O$_4$ nanostructured thin films

Saptarshi De [a], N. Venkataramani [a], Shiva Prasad [b], R. O. Dusane [a], Lionel Presmanes [c], Y. Thimont [c], P. Tailhades [c], Valérie Baco-Carles [c], Corine Bonningue [c], Sumangala T.P [c] and Antoine Barnabé [c]

**Abstract**— Nanocrystalline CuO–CuFe$_2$O$_4$ composite thin films were developed from CuFeO$_2$ ceramic target using a radio frequency sputtering method followed by a thermal oxidation process. This fabrication process helps to develop porous sensing layers which are highly desirable for solid state resistive gas sensors. Their sensing properties towards ethanol and hydrogen gas in dry air were examined at the operating temperatures ranging from 250 °C to 500 °C. The electrical transients during adsorption and desorption of the test gases were fitted with the Langmuir single site gas adsorption model. A composite thin film with a total thickness of 25 nm showed highest response (79%) towards hydrogen (500 ppm) at the operating temperature of 400 °C. The shortest response time ($\tau_{res}$) was found to be ~60 and ~90 seconds for hydrogen and ethanol respectively. The dependence of the response of the sensor on gas concentration (10-500 ppm) was also studied.

## I. INTRODUCTION

Metal oxide semiconductors (MOS), such as pure CuO phase or CuO coupled with other MOS in a composite material, have been used as sensor materials for many years for the detection of reducing gases such as hydrogen [1,2], ethanol [3-7], CO [8,9], and H$_2$S [10-12]. Recently, various nano structures of CuO like one-dimensional (1D) nano wire and thin films have caught attention due to high surface to volume ratio which is expected to enhance the performance of the devices based on semiconductor nano structures [13]. Porous CuO nano wires [14], CuO/ZnO hetero contact sensors [15] and Zn doped CuO nano wires [16] were reported for improving H$_2$ detection. In addition with all the gases listed above, CuO can also be interesting for CO$_2$ detection [17]. On the other hand, copper based spinel oxides such as copper ferrite (CuFe$_2$O$_4$) having n-type semiconductor properties was also reported to show response towards H$_2$ [18], LPG [19] and ethanol [20]. In our previous work, maximum response ($\Delta R/R$) of 86% was obtained by CuFe$_2$O$_4$ nano powder towards 500 ppm of ethanol [21], and this pure copper ferrite also showed a good response of 10% towards CO$_2$ [22].

Semiconductor nano composites with p–n junction were reported as a subject of interest for gas sensing regarding operating temperature (O.T.) and response. In particular, many authors have studied the combination of p-type CuO with various n-type oxides for CO$_2$ detection [23-26]. In the recent past, CuO/CuFe$_2$O$_4$ composite thin films [27] and powders [22] having spinel phase were also reported as CO$_2$ gas sensing material.

In this work, radio-frequency (RF) sputtered CuO/CuFe$_2$O$_4$ semiconductor thin films were used as the sensitive material for reducing gases like hydrogen and ethanol. The effect of the operating temperature on the response, response time and recovery time of the active layer were studied to evaluate the merit of performance of the material. The effect of gas flow rate on the response time and recovery time of the active layer were also studied. To demonstrate its potential sensing application, the variation of response with different gas concentration has been shown. Here, the minimum operating temperature was chosen as 250 °C to avoid the effect of moisture on sensor samples during practical gas sensing application.

## II. PREPARATION OF THE GAS SENSITIVE LAYERS

Cu–Cu$_x$Fe$_{3-x}$O$_4$ thin films were first deposited on fused silica substrate at room temperature with Alcatel A450 RF sputtering unit using a pure delafossite (CuFeO$_2$) ceramic target. The details of the deposition procedure were described by Barnabé *et al.* [28]. Process parameters for the room temperature deposited samples are given in Table I. Two films having thicknesses 25 nm and 300 nm were deposited by varying the deposition time. Thickness of the deposited films was measured using a Dektak 3030ST profilometer and cross-sectional scanning electron microscopy (SEM) using JEOL JSM 6700F field emission SEM. Our previous studies (i.e. grazing incidence X-ray diffraction (GI-XRD), Raman spectroscopy and electron probe micro analysis (EPMA)) on the same samples have already revealed that the as-deposited films consisted of copper nano particles which were embedded in an oxide matrix which was made of cuprous oxide and mixed valence defect ferrite (Cu$_2$O, Cu$_x$Fe$_{3-x}$O$_4$) [28,29]

This work was supported by the MonaSens Project DST (grant number 14IFCPAR001) - ANR (grant number 13-IS08-0002-01). Authors thank IRCC facility, IIT Bombay for the broadband dielectric spectrometer (BDS) facility. One of the authors thanks DST for providing scholarship during this work.

[a] Department of Metallurgical Engineering and Materials Science, Indian Institute of Technology Bombay, Powai, Mumbai 400076, India
[b] Department of Physics, Indian Institute of Technology Bombay, Powai, Mumbai 400076, India
[c] CIRIMAT, Université de Toulouse, CNRS, INPT, UPS, 118 Route de Narbonne, F-31062 Toulouse Cedex 9, France.
e-mail: ramani@iitb.ac.in, sapjaki@gmail.com, barnabe@chimie.ups-tlse.fr

(Equation 1a). In order to obtain the stable CuO/CuFe₂O₄ nano composite, the as-deposited films were ex-situ annealed at 550 °C in air for 12 h (Equation 1b). The tenorite phase CuO then originated from the oxidation of the metallic copper in association with that of Cu₂O. One can note that the reaction scheme could be more complex if we consider the formation of CuFeO₂ intermediate phase [29]. The annealing treatment of the as-deposited samples starting from delafossite target can be represented by the following simplified reaction scheme:

Equation 1a: reduction during deposition step
CuFeO₂→x Cu + (1-x)/2 Cu₂O + (9-x)/24 Fe₂₄/₍₉₋ₓ₎O₄ + x/3 O₂
(Target)           (as-deposited film)
Equation 1b: oxidation during post deposition annealing
x Cu + (1-x)/2 Cu₂O + (9-x)/24 Fe₂₄/₍₉₋ₓ₎O₄+ (4x+3)/12 O₂
(as-deposited film)
→x/2 CuO + x/2 CuFe₂O₄
(annealed film)

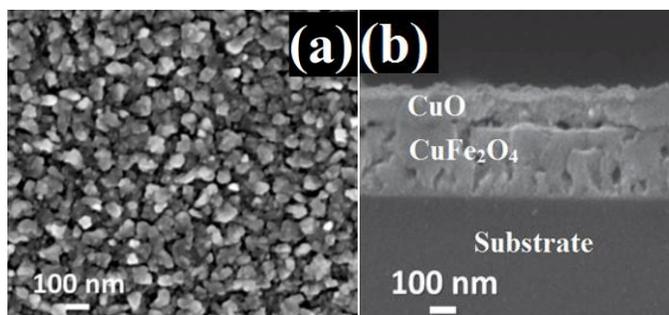

Figure 1: FE-SEM micrographs (a) plain view and (b) cross section view of the sample annealed at 550 °C for 12 hours in air.

The SEM image in figure 1 shows that, as a result of annealing, the parent films were self-organized in a two layered stack with top to bottom layer thickness ratio of 1:2. These films were characterized by GI-XRD technique and Glow-discharge optical emission spectroscopy (GD-OES) profile [27] and X-ray photo electron spectroscopy (XPS) [29] which confirmed that the top layer was tenorite CuO and that of the bottom layer was spinel CuFe₂O₄.

Interestingly, a 30% increase in the total thickness of the as-deposited thin film was observed after annealing which was possibly due to the porosity developed during post-deposition annealing [29]. This porosity in the two layered stack might be caused by the metallic copper diffusion during the oxidation process of the as-deposited samples. For thin film semiconductor metal oxide based gas sensors, the porosity of the sensing layer is an important parameter [30] as the gas diffusion through the porosity can cause changes in electrical properties of the films, making the gas detection easier.

TABLE I
Deposition parameters for the sputtering

| Target | CuFeO₂ |
|---|---|
| Magnetron | No |
| RF power (W) | 200 |
| Argon pressures (Pa) | 0.5 |
| Target to substrate distance (cm) | 5 |
| Substrate | Fused silica |
| Deposition rates (nm/min) | 6.8 |

## III. GAS SENSING MEASUREMENTS

Gas sensing experiments were carried out in a closed chamber with controlled operating temperature from 250 °C to 500 °C using a PID controller. The ambient gas environment was controlled by a continuous flow of the calibrated test gases or air using mass flow controller. For the hydrogen sensing, two gas cylinders were used- one with just zero air (moisture < 0.01%) and another with same zero air containing 500 ppm of hydrogen. The sensor samples were stabilized at each operating temperatures for at least 12 hours in zero air, prior to the gas sensing experiment. Resistance-transients of the sensing layer were measured in two probes mode using Keithley 2635B source meter. Similarly, for the ethanol sensing experiments, two gas cylinders were used, one with zero air and another with the same zero air containing 500 ppm of ethanol. The response ($R_s$) of the sensor samples is defined as the relative difference of the film resistance between air and test gas atmosphere ($R_{gas}$-$R_{air}$) / $R_{air}$ × 100%, where $R_{gas}$ and $R_{air}$ are saturated resistance of the sensor in test gas atmosphere and in air respectively. The concentration of the test gases ($C_{gas\ in\ chamber}$) in the gas chamber was varied by diluting with zero air, and it can be calculated using the following relation:

$$C_{gas\ in\ chamber}=[C_{test\ gas}x(dV_{test}/dt)]/[(dV_{test}/dt)+(dV_{zero\ air}/dt)] \quad (2)$$

Where $C_{test\ gas}$ is the concentration of the test gas in gas cylinder and $dV_{test}/dt$ is the volumetric flow rate of test gas, similarly $dV_{zero\ air}/dt$ is the volumetric flow rate of zero air.

## IV. RESULTS AND DISCUSSION

Figure 2 shows the resistance-transients during the insertion of hydrogen (500 ppm) and recovery in air of the 25 nm thin film sensor at the operating temperature of 400 °C with 100 cc/min gas-flow rate. The sensing material showed good repeatability as the initial baseline was regained upon exposure to dry air. The increase in the electrical resistance of the sensors upon exposure to a reducing gas such as H₂ indicates that the obtained films have p-type semiconducting behaviour. It could be possible that only CuO is involved in sensing as it is the top layer.

The following reaction mechanism for the sensing of reducing gas by a p-type semiconductor can be summarized from several research reports [3]. In a first step, at the operating temperature, oxygen is physisorbed on the sensor surface followed by electron transfer from the p-type semiconducting oxide CuO to the adsorbed oxygen, thus forming chemical bond between the adsorbed oxygen and the semiconducting oxide. Thus, the electrical resistance of the p-type semiconductors reduces during stabilization of the sensor material [see figure (3.a)].

These reactions are described in equations (3) and (4) respectively

$$\tfrac{1}{2}O_2 + [sensor_{surface}] \leftrightarrow O_{ad\text{-}surface} \text{ (physisorption)} \quad (3)$$

$$O_{ad\text{-}surface} + e^- \leftrightarrow O_{ad}^- \text{ (chemisorption)} \quad (4)$$

When the sensor is exposed to reducing gas ambient, the reducing gas is physically adsorbed on the active layer surface and reacts with the adsorbed oxygen according to the reaction (5) & (6) and the product (RO) goes out [eq. 7] [see figure (3.b)]. Thus, the resistance of the p-type sensor increases.

$$R + [\text{sensor}] \leftrightarrow R_{ad} \text{ (physisorption)} \qquad (5)$$
$$R_{ad} + O_{ad}^- \leftrightarrow RO_{ad} + e^- \qquad (6)$$
$$RO_{ad} \leftrightarrow RO_{gas}\uparrow + [\text{sensor}_{\text{ surface}}] \qquad (7)$$

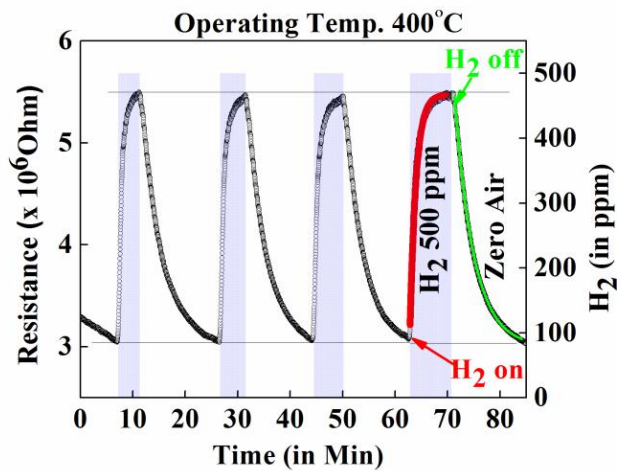

Figure 2: Resistance-transients (response and recovery) of the 25 nm thin film sensor at the operating temperature of 400 °C with 100 cc/min flow rate, fittings are shown in coloured lines.

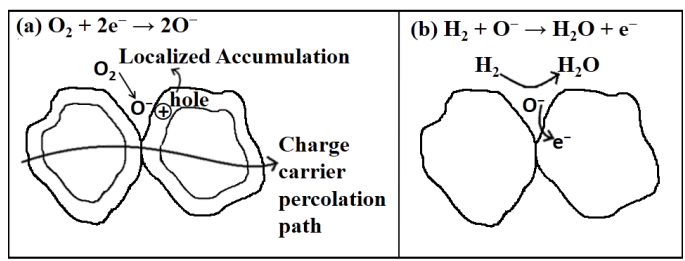

Figure 3: Schematic of the proposed sensing mechanism- (a) during stabilization of the sensor material (oxygen adsorption); (b) during sensing of the test gas (e.g. hydrogen).

Out of these reactions, the physisorption of oxygen as well as that of the reducing gas [Eq. (3) and (5)] are fast. On the other hand the reaction between the adsorbed gas and oxygen [Eq. (6)] is a slow process and therefore, the last one is the rate determining step for the response kinetics. This is easily corroborated from the reported data on surface reaction of adsorbed oxygen [31] and hydrogen [32]. According to Mccoy et al., the ratio of surface reaction rate constant to adsorption rate constant at adsorption equilibrium for oxidation of sulphur dioxide is 0.5 [31], and Arrua et al. reported the same ratio for hydrogenation using Pd/Al₂O₃ catalyst in the range of 0.26-0.29 [32]. Assuming Langmuir single site gas adsorption model for the thin film sensors [33], the response and the recovery transients were fitted well with the following two equations (eq. 8, 9) respectively (shown in coloured lines in fig. 2). The values of coefficient of determination ($R^2$) in this fitting for all response or recovery curves were in between 0.985-0.999.

$$R(t)_{\text{response}} = R_{\text{air}} + R_1[1 - \exp(-t/\tau_{\text{res}})] \qquad (8)$$
$$R(t)_{\text{recovery}} = R_{\text{air}} + R_1[\exp(-t/\tau_{\text{rec}})] \qquad (9)$$

Where $\tau_{\text{res}}$ and $\tau_{\text{rec}}$ are the 'response time' and 'recovery time' respectively. And $R_1$ is a proportionality constant of the exponential term whose value is equal to the difference of the film resistance between air and test gas atmosphere ($R_{\text{gas}} - R_{\text{air}}$).

The variation of response and recovery time with gas flow rate was observed and tabulated in Table II. Decrease in response and recovery time with gas flow rate indicates a mass transfer controlled reaction kinetics on this thin film surface. Therefore, the flow rate was kept fixed at 100 cc/min for the rest of the experiments performed.

TABLE II
Variation in response time and recovery time with gas-flow rate at a fixed operating temperature (500 °C); 25 nm composite thin film

| Gas flow rate (in cc/min) | $\tau_{\text{res}}$ (in s) | $\tau_{\text{rec}}$ (in s) |
|---|---|---|
| 20 | 137 | 221 |
| 50 | 82 | 190 |
| 100 | 63 | 131 |

Hydrogen sensing by a 25 nm thin film sensor was carried out at different operating temperatures and the variation of response time and recovery time are given in the table III. Response time seemed to be saturated above the operating temperature of 350 °C and the saturated value was found to be around 60 seconds. At the high operating temperatures, the reaction rate of eq. (6) became faster and the reaction might be limited by the test conditions, i.e., gas flow in the gas chamber. Recovery time decreased monotonically with the operating temperature until 500 °C.

TABLE III
Variation in response time and recovery time of 25 nm thin film sensor with the operating temperature (test gas: 500 ppm of H₂ with 100 cc/min flow rate)

| Operating Temperature (in °C) | $\tau_{\text{res}}$ (in s) | $\tau_{\text{rec}}$ (in s) |
|---|---|---|
| 250 | 276 | 667 |
| 300 | 116 | 416 |
| 350 | 58 | 249 |
| 400 | 58 | 221 |
| 450 | 56 | 186 |
| 500 | 63 | 131 |

The bell shaped response curve with the operating temperature as shown in the figure 4 is a result of the competitive behaviour of eqs. (5) and (6). Similar bell shaped response curve had been reported by Ahlers et al. [34] and Biswas et al. [35]. According to them, response varies with operating temperature on the basis of two energy systems. $E_{ads}$ is dependent on the strength of the test gas binding onto the sensing material surface. On the other hand, $E_a$ is defined as the energy barrier required to be overcome by the adsorbed gas molecules for diffusion along the surface, resulting in catalysis induced surface combustion process. Initially, under clean air conditions, active sites on the surface of a sensor material had been covered by adsorbed oxygen. Then, relative occupancy of the test gas on the pre-adsorbed oxygen depends on partial pressure and operating temperature of the test gas.

In the figure 4, simulated curves of Langmuir relative surface coverage (L), reaction rate (K) and the modelled response ($R_m$) are shown.

$$L = \frac{P_{gas}}{P_{gas} + P_o} \qquad (10)$$

where $P_{gas}$ is partial pressure of test gas ($H_2$) at sensing layer and $P_o = \frac{k_B T}{V_o} \exp\left(\frac{-E_{ads}}{k_B T}\right)$, where $v_o$ is the quantum volume of the test gas species, given by $v_o = \left(\frac{2\pi\hbar^2}{M_{gas} M_o k_B T}\right)^{1.5}$, where $M_{gas}$ is the relative atomic mass of the test gas (i.e. 2 for $H_2$); $M_o$ is the atomic mass unit ($1.67\times10^{-27}$ kg); $\hbar$ is the reduced plank constant and $k_B$, T are Boltzmann constant and absolute temperature respectively. The reaction rate of adsorbed test gas with chemisorbed oxygen ion is

$$K = A \exp\left(\frac{-E_a}{k_B T}\right) \qquad (11)$$

where A is a proportionality constant. The modelled response was obtained from the combination of Langmuir relative surface coverage and the reaction rate [34], and it could be given by

$$R_m = \frac{P_{gas}}{P_{gas} + \frac{k_B T}{V_o} \exp\left(\frac{-E_{ads}}{k_B T}\right)} A \exp\left(\frac{-E_a}{k_B T}\right) \qquad (12)$$

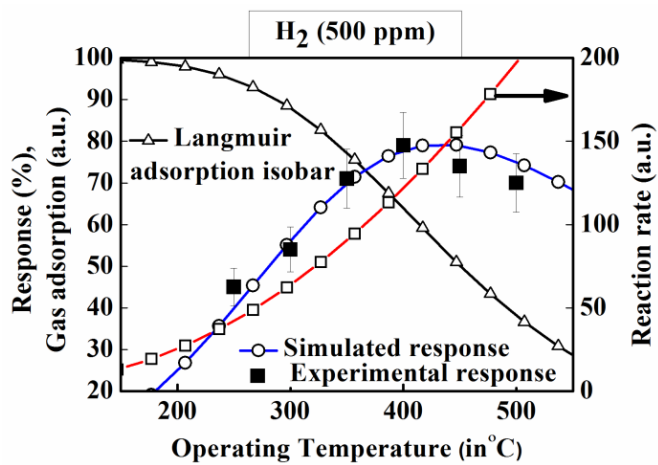

Figure 4: Experimental and simulated response vs. operating temperature of 25 nm composite thin film (test gas: 500 ppm of $H_2$ with 100 cc/min flow rate).

In the case of tin dioxide ($SnO_2$) thin film sensors, the values of $E_{ads}$ varied from 130 to 145 kJ/mol (1.3-1.45 eV) and $E_a$ varied from 53 to 57 kJ/mol (0.53-0.57 eV) for different ethane concentrations [34]. Here, the values $E_{ads}$ and $E_a$ were obtained (by the fitting of experimental values with eq. 12) as 43 kJ/mol (0.45 eV) and 21 kJ/mol (0.22 eV) respectively. This sensor sample showed maximum response of 79% at the operating temperature of 400 °C towards 500 ppm of $H_2$. A similar response of 70% was reported for $H_2$ but at a higher concentration (2500 ppm) with thicker copper oxide –copper ferrite sensor system [36]. N.D. Hoa et al. reported 40% response towards 10,000 ppm of $H_2$ at an operating temperature of 250 °C for CuO thin film, whereas at the same operating temperature, this CuO/$CuFe_2O_4$ thin film exhibits 45% response only at 500 ppm of $H_2$ [2].

TABLE IV
Change in response time and recovery time of 25 nm thin film sensing material with the operating temperature (test gas: 500 ppm of ethanol with 100 cc/min flow rate)

| Operating Temperature (in °C) | $\tau_{res}$ (in s) | $\tau_{rec}$ (in s) |
|---|---|---|
| 250 | 432 | 740 |
| 300 | 223 | 440 |
| 350 | 150 | 370 |
| 400 | 121 | 390 |
| 450 | 90 | 450 |
| 500 | 90 | 480 |

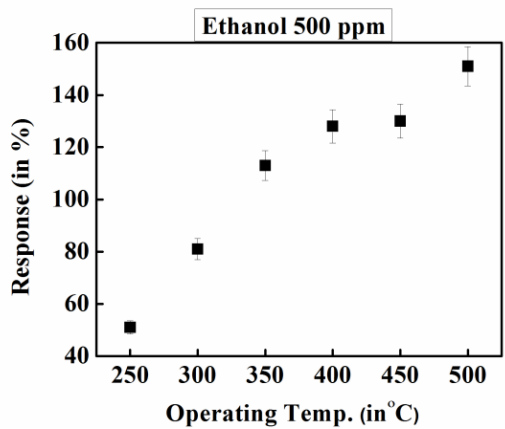

Figure 5: Response vs. operating temperature of 25 nm thin film sensing material (test gas: 500 ppm of ethanol with 100 cc/min flow rate).

Similarly, ethanol sensing of the 25 nm thin film sensor was carried out at different operating temperatures and the variation of response time and recovery time are given in the table IV. Response time seemed to be saturated above the operating temperature of 450 °C and the value of that was found to be around 90 seconds. It was observed that the recovery time decreased to the range of 350 to 400 °C and after that it had increased. This increase at 500 °C is not well understood at present. Figure 5 shows the variation of response with operating temperature. In case of ethanol, the maximum response of this thin film sensor might be observed above 500 °C. As per literature, the best operating temperature to get maximum response for ethanol was reported to be higher than that of hydrogen [37]. The operating temperature was confined below 500 °C due to the (a) stability of the phase and microstructure in sensing layers and (b) instrumental limitation. Hence we may not be able to capture a similar behaviour as found in $H_2$. Here, due to lack of the bell shape in the response curve, it could not be fitted with the chosen model (eq.12).

TABLE V
Response time, recovery time and response vs. film-thickness

| Thickness of films (nm) | Test gas (500 ppm) | Operating Temperature (°C) | $\tau_{res}$ (s) | $\tau_{rec}$ (s) | $R_s$ (%) |
|---|---|---|---|---|---|
| 25 | Ethanol | 450 | 90 | 450 | 130 |
| 300 | Ethanol | 450 | 204 | 800 | 114 |
| 25 | $H_2$ | 400 | 58 | 221 | 79 |
| 300 | $H_2$ | 400 | 230 | 790 | 68 |

The gas sensing experiment was carried out with 300 nm thick film and the results are tabulated in table V. Thicker $CuO–CuFe_2O_4$ thin film showed p-type response towards reducing gases, i.e., hydrogen and ethanol. So, the test gases were mostly interacting (adsorption / desorption) with the copper oxide layer located on the top of film.

Depending on the operating temperature, oxygen molecules adsorbed on semiconductor surface are in various ionic states, i.e. $O_2^-$, $O^-$ or $O^{2-}$ [38]. So, adsorbed hydrogen ($H_{2ads}$) may react with adsorbed oxygen ($O^{ion}_{ads}$) as in the following equations.

$$2H_{2\,ads} + O_2^-{}_{ads} \rightarrow 2H_2O + e^- \qquad (13)$$
or,
$$H_{2\,ads} + O^-{}_{ads} \rightarrow H_2O + e^- \qquad (14)$$
or,
$$H_{2\,ads} + O^{2-}{}_{ads} \rightarrow H_2O + 2e^- \qquad (15)$$

Similarly, carbon dioxide and water are the final decomposition products of ethanol combustion in air. Acetaldehyde or acetic acid may also form as intermediate products during the oxidization of ethanol. Hence depending on the types of adsorbed oxygen and by-products of ethanol, various charge balance equations of ethanol decomposition are possible and given below.

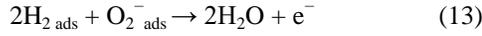
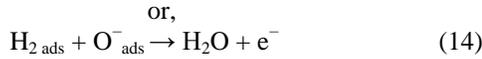
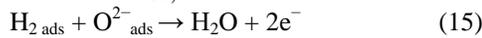

$$
\left.
\begin{aligned}
&2CH_3CH_2OH_{ads} + O_2^-{}_{ads} \rightarrow 2CH_3CHO + 2H_2O + e^-\\
&CH_3CH_2OH_{ads} + O^-{}_{ads} \rightarrow CH_3CHO + H_2O + e^-\\
&CH_3CH_2OH_{ads} + O^{2-}{}_{ads} \rightarrow CH_3CHO + H_2O + 2e^-\\
&CH_3CH_2OH_{ads} + O_2^-{}_{ads} \rightarrow CH_3COOH + H_2O + 1e^-\\
&CH_3CH_2OH_{ads} + 2O^-{}_{ads} \rightarrow CH_3COOH + H_2O + 2e^-\\
&CH_3CH_2OH_{ads} + 2O^{2-}{}_{ads} \rightarrow CH_3COOH + H_2O + 4e^-\\
&CH_3CH_2OH_{ads} + 3O_2^-{}_{ads} \rightarrow 2CO_2 + 3H_2O + 3e^-\\
&CH_3CH_2OH_{ads} + 6O^-{}_{ads} \rightarrow 2CO_2 + 3H_2O + 6e^-\\
&CH_3CH_2OH_{ads} + 6O^{2-}{}_{ads} \rightarrow 2CO_2 + 3H_2O + 12e^-
\end{aligned}
\right\} (16)
$$

From equations (13), (14) and (15), for all metal oxide sensors a general rate equation of electron density can be written as

$$\frac{dn}{dt} = K_{gas}(T)[O^{ion}_{ads}]^a [R]^b \qquad (17)$$

where, n is the electron density or electron concentration in the charge accumulation layer under the test gas (e.g. $H_2$) atmosphere, b is a charge parameter which might have value in the range of 0.5 to 2 for hydrogen and 0.08 to 2 for ethanol respectively. Similarly, a is a charge parameter which might have value in the range of 0.5 to 1 for oxygen ions. $K_{gas}$ (T) is the reaction rate constant or reaction rate coefficient described as

$$K_{gas}(T) = A \exp(-E_a/k_BT) \qquad (18)$$

where $E_a$ is the activation energy of reaction, $k_B$ is the Boltzmann constant, T is absolute temperature and A is proportionality constant. Integrating Eq. (17) leads to the solution as

$$n = K_{gas}(T)[O^{ion}_{ads}]^a [R]^b t + n_o \qquad (19)$$

where $n_o$ is the saturated electron concentration of sensor at an operating temperature in the air atmosphere. In the saturated ethanol, i.e., at equilibrium under ethanol and air atmosphere, carrier concentration n and $n_o$ could be considered as a constant with time.

$$n = K_{gas}(T)[O^{ion}_{ads}]^a [R]^b \tau + n_o \qquad (20)$$

Where $\tau$ is a time constant. At a constant operating temperature the resistivity of a semiconductor is defined as $\rho = \alpha / n$. Where $\alpha$ is a proportionality constant with '+' sign for n-type and '–' sign for p-type semiconductor, and can be substituted in equation (20) as

$$\frac{1}{Rg} = (K_{gas}(T)[O^{ion}_{ads}]^a [R]^b \tau) / \alpha + \frac{1}{Ra} \qquad (21)$$

Assuming the concentration of adsorbed test gas ($[R]^b$) on the sensor surface is linearly proportional to the gas concentration in gas chamber ($C_g^b$), at constant operating temperature the sensor response relation can be obtained in a compact form

$$R_s = MC_g^b \qquad (22)$$

where $R_s$ is response of the sensor and it could be defined as $(R_{gas}-R_{air})/R_{air}$ and M is $(K_{gas}(T)[O^{ion}_{ads}]^a \tau)R_{air} / \alpha$, a constant at constant operating temperature.

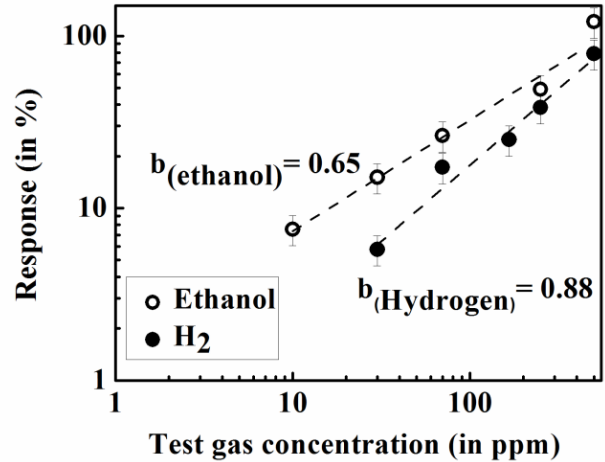

Figure 6: Response vs. test gas concentration of 25 nm composite thin film at the operating temperature of 400 ℃; gas flow rate: 100 cc/min.

Figure 6 shows the variation of response of the 25 nm thin film sensor with gas concentration at the operating temperature of 400 ℃. Gas sensing response is following the power law equation (eq. 22) for both the gases in the range of 10 ppm to 500 ppm. Response towards ethanol is slightly higher than that of hydrogen for similar concentration, i.e., this sensor is more sensitive towards ethanol than hydrogen. The obtained value of b is 0.65 for ethanol and 0.88 for hydrogen. This value of b towards ethanol is quite similar with the reported values, 0.677 and 0.54 for ZnO nano rods and nano structured sensing materials respectively [39,40]. For hydrogen, the value of b was reported as 0.53 for ZnO thin films [41]. The value of b of these sensors was not as close to 0.5. Such deviation might occur due to the fact that the surface depletion or accumulation layer has some effect on the oxygen adsorption species at metal oxide surface when the grain diameter is close to double of that layer thickness ($2L_d$) [41].

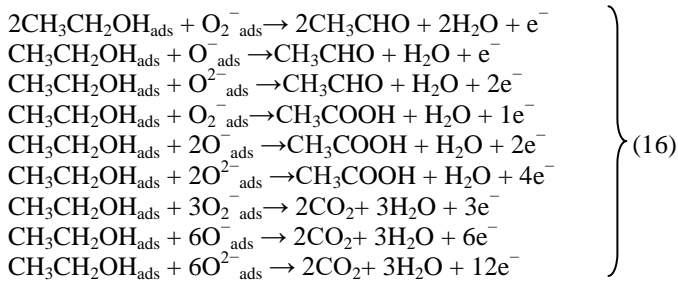

Comparison of our experimental data with available literature values of CuO based sensing materials

| Sensor material | Test gas (ppm) | $\tau_{res}$ | $\tau_{rec}$ | $R_s$ (%) | O.T. (°C) | Ref. |
|---|---|---|---|---|---|---|
| CuO thin film | $H_2$ (10,000) | ~10 min | ~20 min | 40 | 250 | [2] |
| CuO nanostructures | $H_2$(100) | - | - | 50 | 300 | [42] |
| Porous CuO nanowires | $H_2$(60,000) | ~2.5 min | ~10 min | 400 | 250 | [14] |
| CuO/ZnO hetero contact | $H_2$(4000) | - | - | 25 | 200 | [1] |
| CuO–CuFe$_2$O$_4$ thin film | $H_2$(1250) $H_2$(2500) | 190 s - | 400 s - | 40 79 | 295 295 | [36] |
| CuO/ZnO hetero contacts | $H_2$(4000) | - | - | 130 | 400 | [15] |
| CuO–CuFe$_2$O$_4$ thin film | $H_2$(500) | 58 s | 221 s | 79 | 400 | our work |
| CuO nano rods | Ethanol (2000) | - | - | 160 | 300 | [3] |
| CuO nano wires | Ethanol (500) | - | - | 24 | 300 | [4] |
| CuO nano wires | Ethanol (1000) | 110 s | 120 s | 50 | 240 | [5] |
| CuO microspheres | Ethanol (200) | - | - | 70 | 240 | [6] |
| CuO thin film | Ethanol (12.5) | - | - | 120 | 180 | [7] |
| CuO–CuFe$_2$O$_4$ thin film | Ethanol (500) | 90 s | 450 s | 130 | 450 | our work |

The ethanol and $H_2$ sensing properties of various CuO nano structures in the literature are summarized in Table VI. Few of them reported higher response in comparison to the current work but at the cost of very high gas concentration [3,14-15]. And short response time was observed in this present study among the values reported recently in the literature of CuO sensors.

## V. CONCLUSION

The self-organized CuO–CuFe$_2$O$_4$ thin films showed p-type semiconductor behaviour with increase in electrical resistance upon exposure to hydrogen or ethanol gas. Good fitting of response or recovery curve with single site gas adsorption model indicates that the reaction had occurred only on the surface of thin films. The developing process of this porous microstructure of top CuO layer is interesting as this kind of sensors have shown improved sensing properties compared to the CuO thin film sensors fabricated by other techniques already reported. The best sensing performance was observed for the 25 nm thin film at an operating temperature of 400 °C with a response of 79% towards 500 ppm of $H_2$ and the response and recovery times obtained at this temperature are ~60 s and ~220 s, respectively. This 25 nm thin film sample also exhibited 128% response towards 500 ppm of ethanol with 90 seconds response time at the operating temperature of 400 °C. Also, we have demonstrated the variation of response of the sensors for a wide range of test-gas concentration. Due to these promising results, we believe that an optimised fabrication of gas sensors made from this composite material could be a potential candidate for the cheap hydrogen leak detectors, breathalyzer and other similar applications.